\documentclass[preprint,12pt]{elsarticle}

\usepackage{epsfig}

\usepackage{amssymb}

\journal{Physics Letters B}

\begin{document}

\begin{frontmatter}



\title{Radiative viscosity of neutron stars}



\author{Shu-Hua Yang\corref{cor1}}
\ead{ysh@phy.ccnu.edu.cn}
\author{Xiao-Ping Zheng}
\author{Chun-Mei Pi}
\address{Institute of Astrophysics, Huazhong Normal University,  Wuhan 430079, China}

\begin{abstract}
We study non-linear effects of radiative viscosity of $npe$ matter in neutron stars
for both direct Urca process and modified Urca process, and find that
non-linear effects will decrease the ratio of radiative viscosity to
bulk viscosity from $1.5$ to 0.5 (for direct Urca process)
and 0.375 (for modified Urca process). Which means that for small
oscillations of neutron star, the large fraction of oscillation energy is
emitted as neutrinos; but for large enough ones, bulk viscous
dissipation dominates.
\end{abstract}


\end{frontmatter}


\section{Introduction}
\label{1}
Bulk viscosity can damp the density oscillations in compact stars, which
could be excited at the time of their formation from supernovae explosions,
or during the phase transitions \cite{mig79}, or due to instabilities
result from the emission of gravitational waves \cite{and98, fri98, lin98, mad00, and01}.
As one of the most important transport coefficients,
bulk viscosities of simple $npe$ matter, of hyperon matter and  even of quark matter,
both in normal and superfuild states, have been extensively studied
\cite{fin68, saw8901, haensel92, hae01, wan84, saw8902, mad92, gup97, lin02, zhe02, zhe04, zhe05, pan06, sad0701, sad0702},
for more references see \cite{don07}. However, until recently, Sa'd et al. \cite{sad09}
demonstrated that there exists a new mechanism for dissipating the
energy of stellar oscillations. They indicated that the mechanical
energy of density perturbations is not only disspiated to
heat via bulk viscosity, but also is radiated away
via neutrinos. They named this new mechanism the radiative viscosity,
and found it is 1.5 times larger than the bulk viscosity to all
Urca processes, both in nuclear matter and quark matter.

The newly realized radiative viscosity was calculated only in
the lowest order of $\delta \mu/ T$ \cite{sad09}, which corresponds to
the linear approximation. However, the density oscillations during
the stellar evolution may arise to sufficiently large amplitude so
that no-linear effects can no longer be neglected. We
aim to study the non-linear effects of radiative viscosity in this paper.

We will study non-linear effects of radiative viscosity of simple $npe$ matter,
both for direct Urca process ($n\rightarrow p+e+\overline{\nu}_{e}$, $p+e\rightarrow n+\nu_{e}$)
and for modified Urca process ($n+N\rightarrow p+N+e+\overline{\nu}_{e}$, $p+N+e\rightarrow n+N+\nu_{e}$).
As indicated by Lattimer et al. \cite{lat91},
if the proton and electron Fermi momenta are
too small compared with neutron Fermi momenta, the nucleon direct
Urca process is forbidden because it is impossible to satisfy conservation of
momentum. Under typical conditions, one finds that the ratio of the
number density of protons to that of nucleons must exceed
1/9 for the process to be allowed.

We not only make a numerical solution of radiative viscosity bue also study the ratio
of the viscosity coefficient to the bulk one. By doing so, we hope to know which mechanism is
more important under special condition.

This paper is arranged as follows. In Sect.2, we derive the formulae of radiative viscosity including
non-linear effects. And radiative viscosity for a specific model of neutron star (NS) matter
is calculated in Sect.3. Finally, Sect.4 presents the conclusions and discussions.

\section{The formulae of radiative viscosity}
\label{2}

In this section we derive the expression for radiative viscosity of neutron
stars whose cores consist of simple $npe$ matter using an approach similar to that of
bulk viscosity calculations has been done by
Wang and Lu \cite{wan84}, Sawyer \cite{saw8902}, Madsen \cite{mad92} and Gupta et al. \cite{gup97}.

We first recall how to solve chemical potential perturbations with stellar oscillations.
Assume that the volume per unit mass, $\upsilon$, changes periodically in time according
to the relation
\begin{equation}
\upsilon(t)= \upsilon_{0}+\Delta \upsilon {\rm sin} \left( \frac{2\pi t}{\tau}\right)
=\upsilon_{0}+\delta \upsilon(t),
\end{equation}
where $\upsilon_{0}$ is the equilibrium volume, $\Delta \upsilon$ is the perturbation
amplitude, and $\tau$ is the period.

Taking the chemical potential difference as
\begin{equation}
\delta \mu \equiv \mu_{p}+\mu_{e}-\mu_{n},
\end{equation}
where $\mu_{n}$, $\mu_{p}$ and $\mu_{e}$ are the chemical
potentials of the neutrons, protons and electrons.
$\delta \mu$ can be expanded near the equilibrium according to
\begin{equation}
\delta \mu (t)= \left( \frac{\partial \delta \mu}{\partial \upsilon}\right)_{0}\delta \upsilon
+ \left( \frac{\partial \delta \mu}{\partial n_{p}}\right)_{0} \delta n_{p}
+ \left( \frac{\partial \delta \mu}{\partial n_{n}}\right)_{0} \delta n_{n}
+ \left( \frac{\partial \delta \mu}{\partial n_{e}}\right)_{0} \delta n_{e},
\end{equation}
(note that $\delta \mu=0$ in equilibrium) $n_{i}$ are particle numbers per unit mass and
\begin{equation}
\delta n_{p}=-\delta n_{n}=\delta n_{e}=\int_{0}^{t}(dn_{p}/dt)dt.
\end{equation}
Using the thermodynamical relations
\begin{equation}
\frac{\partial \mu_{i}}{\partial \upsilon}=-\frac{\partial P}{\partial n_{i}},
\end{equation}
and considering $\rho_{i}=n_{i}/\upsilon_{0}$
($\rho_{i}$ is the particle numbers per unit volume), one gets
\begin{equation}
\delta \mu (t)
=-A \frac{\Delta\upsilon}{\upsilon_{0}}{\rm sin} \left( \frac{2\pi t}{\tau}\right)
+B \int_{0}^{t}(d\rho_{p}/dt)dt \label{delmu},
\end{equation}
where
\begin{eqnarray}
A=\left( \frac{\partial P}{\partial \rho_{p}}\right)_{0}
-\left( \frac{\partial P}{\partial \rho_{n}}\right)_{0}
+\left( \frac{\partial P}{\partial \rho_{e}}\right)_{0},\label{a0} \\
B=\left( \frac{\partial \delta \mu}{\partial \rho_{p}}\right)_{0}
-\left( \frac{\partial \delta \mu}{\partial \rho_{n}}\right)_{0}
+\left( \frac{\partial \delta \mu}{\partial \rho_{e}}\right)_{0}\label{b0},
\end{eqnarray}
and $P(t)$ is the pressure,
and the net reaction rate is
\begin{equation}
\frac{d\rho_{p}}{dt}=\Gamma_{\overline{\nu}}-\Gamma_{\nu}=-\Gamma(T,\delta\mu)\label{rout}.
\end{equation}
According to \cite{rei95, hae92} and introducing $z=\delta\mu/\pi T$, for direct Urca process
\begin{equation}
\Gamma_{d}(T,\delta\mu)\delta\mu
=\epsilon_{d}(T,0)\frac{714z^{2}+420z^{4}+42z^{6}}{457} \label{gammad},
\end{equation}
\begin{equation}
\epsilon_{d}(T,0)=3.3 \times 10^{-14} \left(\frac{x_{p}\rho}{\rho_{0}} \right)^{1/3}T^{6} {\rm MeV^{5}} ,
\end{equation}
and for modified Urca process
\begin{equation}
\Gamma_{m}(T,\delta\mu)\delta\mu
=\epsilon_{m}(T,0) \frac{14,680z^{2}+7560z^{4}+840z^{6}+24z^{8}}{11,513} \label{gammam},
\end{equation}

\begin{equation}
\epsilon_{m}(T,0)=3.6 \times 10^{-18} \left(\frac{x_{p}\rho}{\rho_{0}}\right)^{1/3}T^{8} {\rm MeV^{5}},
\end{equation}
where $x_{p}=\rho_{p}/\rho_{b}$ is the ratio of the
number density of protons to that of nucleons, and $\rho_{0}=0.16fm^{-3}$.

Given a specific equation of state (EOS) of $npe$ matter, $\delta\mu(t)$ can be calculated from eqs. ($\ref{delmu}$), ($\ref{a0}$), ($\ref{b0}$) and ($\ref{rout}$) numerically, which will
be done in section 3. Once the time dependence of chemical potential difference is known,
one can easily calculate the bulk and radiative viscous coefficient.

For a periodic process, the expansion and contraction of the system
will induce not only the dissipation of oscillation energy to
heat, but also the loss of oscillation energy through
an increasing of the neutrino emissivity.
Bulk viscous coefficient $\zeta$ and radiative viscous coefficient ${\cal R}$ can be
defined for the description of these dissipation mechanisms, respectively \cite{sad09}
\begin {equation}
\langle \dot{\cal E}_{\rm diss}\rangle =-\frac{\zeta}{\tau}
\int_0^{\tau} dt \left(\nabla \cdot \vec v\right)^2,
\label{epsilon-kin}
\end{equation}

\begin {equation}
\langle \dot{\cal E}_{\rm loss}\rangle =\frac{{\cal R}}{\tau}
\int_0^{\tau} dt \left(\nabla \cdot \vec v\right)^2,
\label{epsilon-kin}
\end{equation}
where $\vec v$ is the hydrodynamic velocity associated with the density oscillations.

Using the continuity equation, one obtains
\begin{equation}
\zeta=-2 \langle \dot{\cal E}_{\rm diss}\rangle
\left( \frac{\upsilon_0}{\Delta \upsilon } \right)^{2} \left( \frac{\tau}{2\pi}\right)^{2} \label{z},
\end{equation}
\begin{equation}
{\cal R}=2 \langle \dot{\cal E}_{\rm loss} \rangle
\left( \frac{\upsilon_0}{\Delta \upsilon } \right)^{2} \left( \frac{\tau}{2\pi}\right)^{2} \label{r},
\end{equation}
where the energy dissipation is
\begin{equation}
 \langle \dot{\cal E}_{\rm diss}\rangle=-\int_{0}^{\tau} \Gamma(T,\delta\mu)\delta\mu dt,
\end{equation}
and the neutrino emissivity
caused by the oscillation of $\delta\mu$ is
\begin{equation}
\langle  \dot{\cal E}_{\rm loss}  \rangle=\int_{0}^{\tau} \left[\epsilon(T,\delta\mu)-\epsilon(T,0)\right] dt,
\end{equation}
for direct Urca process \cite{rei95}
\begin{equation}
\epsilon_{d}(T,\delta\mu)=\epsilon_{d}(T,0)\left(1+\frac{1071z^{2}+315z^{4}+21z^{6}}{457}\right) \label{epsilond},
\end{equation}
and for modified Urca process
\begin{equation}
\epsilon_{m}(T,\delta\mu)=\epsilon_{m}(T,0)
\left(1+\frac{22,020z^{2}+5640z^{4}+420z^{6}+9z^{8}}{11,513}\right)\label{epsilonm}.
\end{equation}

\section{Radiative viscosity for a specific model of NS matter}
\label{3}
\subsection{EOS model}
For illustration, we use a phenomenological EOS proposed by Prakash et al. \cite{pra88}.
According to these authors, the nuclear energy is presented in the form
\begin{equation}
E_{N}(\rho_{b},x_{p})=E_{N0}(\rho_{b})+S(\rho_{b})(1-2x_{p})^{2},
\end{equation}
where $E_{N}(\rho_{b},x_{p})=E_{N0}(\rho_{b})(\rho_{b},x_{p}=1/2)$
is the energy of symmetric nuclear matter and $S(\rho_{b})$ is the
symmetry energy. Supposing the electron energy is $E_{e}(\rho_{b},x_{p})$,
one has $\partial [E_{N}(\rho_{b},x_{p}) + E_{e}(\rho_{b},x_{p})]/\partial x_{p}=0$
in $\beta$ equilibrium.

At the saturation density $\rho_{0}=0.16fm^{-3}$ the symmetry
energy is measured in laboratory, $S(\rho_{0})=30 \rm {MeV}$; while at higher
$\rho_{b}$ it is still unknown.  Prakash et al. \cite{pra88} displayed
$S(\rho_{b})$ in the following form
\begin{equation}
S(\rho_{b})= 13 {\rm {MeV}} \left[u^{2/3}-F(u)\right]+S(\rho_{0})F(u),
\end{equation}
where $u=\rho_{b}/\rho_{0}$ and $F(u)$ satifies the condition $F(1)=1$.
We employ their model I ($F(u)=u$) with the compression modulus of
saturated nuclear matter $K=240$MeV. This is a moderately stiff EOS, the maximum stellar mass
for it is $M=1.977M_{\odot}$. The direct Urca process is allowed only
if $x_{p}>1/9$; in our EOS model it demands $\rho_{b}>0.434 {\rm {fm^{-3}}}$,
which means the direct Urca process is completely forbidden in NS
when its mass is smaller than $M_{D}=1.358M_{\odot}$.

Considering $\rho_{p}=\rho_{e}=\rho_{b}x_{p}$ and $\rho_{n}=\rho_{b}(1-x_{p})$, one has
\begin{eqnarray}
A
=\frac{1}{\rho_{b}}\left( \frac{\partial P}{\partial x_{p}}\right)_{0},\\
B
=\frac{1}{\rho_{b}}\left( \frac{\partial \delta \mu}{\partial x_{p}}\right)_{0}.
\end{eqnarray}
Thus,  $\delta\mu(t)$ can be solved numerically,
and the bulk and radiative viscosity can be calculated from eqs. ($\ref{z}$) and ($\ref{r}$), respectively.

\subsection{The results}
The solid line in Fig.1 and Fig.2 shows radiative viscosity
as function of relative volume perturbation amplitude for
$\rho_{b}=0.6fm^{-3}$ and $\rho_{b}=0.3fm^{-3}$, note that the onset of the direct Urca process
is $\rho_{b}=0.434$. As indicated in ref. \cite{gup97}, the flat branches in Fig.1 for
the direct Urca process follows the $T^{4}$
law: when $T$ increases by four orders (form $10^{-4}$MeV to $1$MeV), viscosity increases by about 16 orders.
In contrast, the flat branches in Fig.2 for the modified Urca process follows the $T^{6}$ law, and
the radiative viscosity of modified Urca process is much lower than direct Urca process.
However, the behavior of the onset of non-linear effect is similar for the two different processes:
the lower the temperature, the smaller the relative volume perturbation amplitude is for
non-linear effects to dominate. In the other words, as $T$ decreases, non-linear effect
becomes more and more important. Further more, we also present the results up to $z^{4}$.
As we all know, when calculated up to $z^{2}$, one can only get the linear results; and
non-linear effects are usually introduced by calculating up to $z^{4}$. But
by comparing solid and dash curves in Fig.1 and Fig.2, we can see that it's not
enough just including the lowest order of non-linear effects.

Fig.3 presents the ratio of radiative viscosity to bulk viscosity as
function of relative volume perturbation amplitude. For both direct Urca
and modified Urca, ${\cal R}/\zeta=1.5$ when $\Delta \upsilon/\upsilon_{0}$
is sufficiently small, which is in agreement with the results given by
Sa'd and Schaffner-Bielich \cite{sad09}. And for both processes,
non-linear effects lead to the decreasing of ${\cal R}/\zeta$, and the lower
the temperature, the more remarkable the non-linear effect is.
Nevertheless, for direct Urca process, ${\cal R}/\zeta$ can reach
0.5 when $\Delta \upsilon/\upsilon_{0}$ is large enough; but
${\cal R}/\zeta$ can be equivalent to 0.375 for modified Urca process.
Comparing the formulae ($\ref{gammad}$) with ($\ref{epsilond}$),  and ($\ref{gammam}$) with ($\ref{epsilonm}$),
one immediately know that the ratios
only depend on the average value of the net reaction rate and the
increments of neutrino emissivity due to off-equilibrium, and EOS can't
effect the results at all.
It is also
interesting to plot ${\cal R}/\zeta$ as function of
$\Delta\mu/T$ (Fig.4), where $\Delta\mu$ is the perturbation
amplitude of chemical potential difference $\delta\mu(t)$.
It's undoubtable that the curves are overlapped for
different values of temperature in Fig.4.

\section{Conclusions and discussions}
\label{4}
We have studied the non-linear effects of radiative viscosity of simple $npe$ matter,
both for direct Urca process and for modified Urca process, and found that
non-linear effects decrease ${\cal R}/\zeta$ from 1.5 (the linear scenario)
to 0.5 (for direct Urca process) and 0.375 (for modified Urca process).
Which means that in the linear scenario, only $\frac{2}{5}$ times
of the total oscillation energy is converted to heat and the
large faction of the energy is emitted as neutrinos; in contrast,
if the amplitude of the oscillation is large enough and
the non-linear effects cannot be ignored, the main
part of the total oscillation energy is converted to heat (
$\frac{2}{3}$ for direct Urca process and $\frac{8}{11}$ for modified process).
Although the numerical results of the coefficients are calculated for
a specific EOS, the ratio of the two types of viscosity coefficient
is EOS-independent.

In the case of superfluid $npe$ matter, since superfluidity
causes different effects to the net reaction rate and
the increments of neutrino emissivity due to off-equilibrium,
we expect that both in linear regime and in non-linear regime, the
value of ${\cal R}/\zeta$ will have great differences comparing with
the normal EOS. This is the further work we set about.

\section*{Acknowledgments}
This research was supported by NFSC under Grants No. 10773004.


\clearpage
 \begin{figure}
   \centering
   \includegraphics[width=0.5\textwidth]{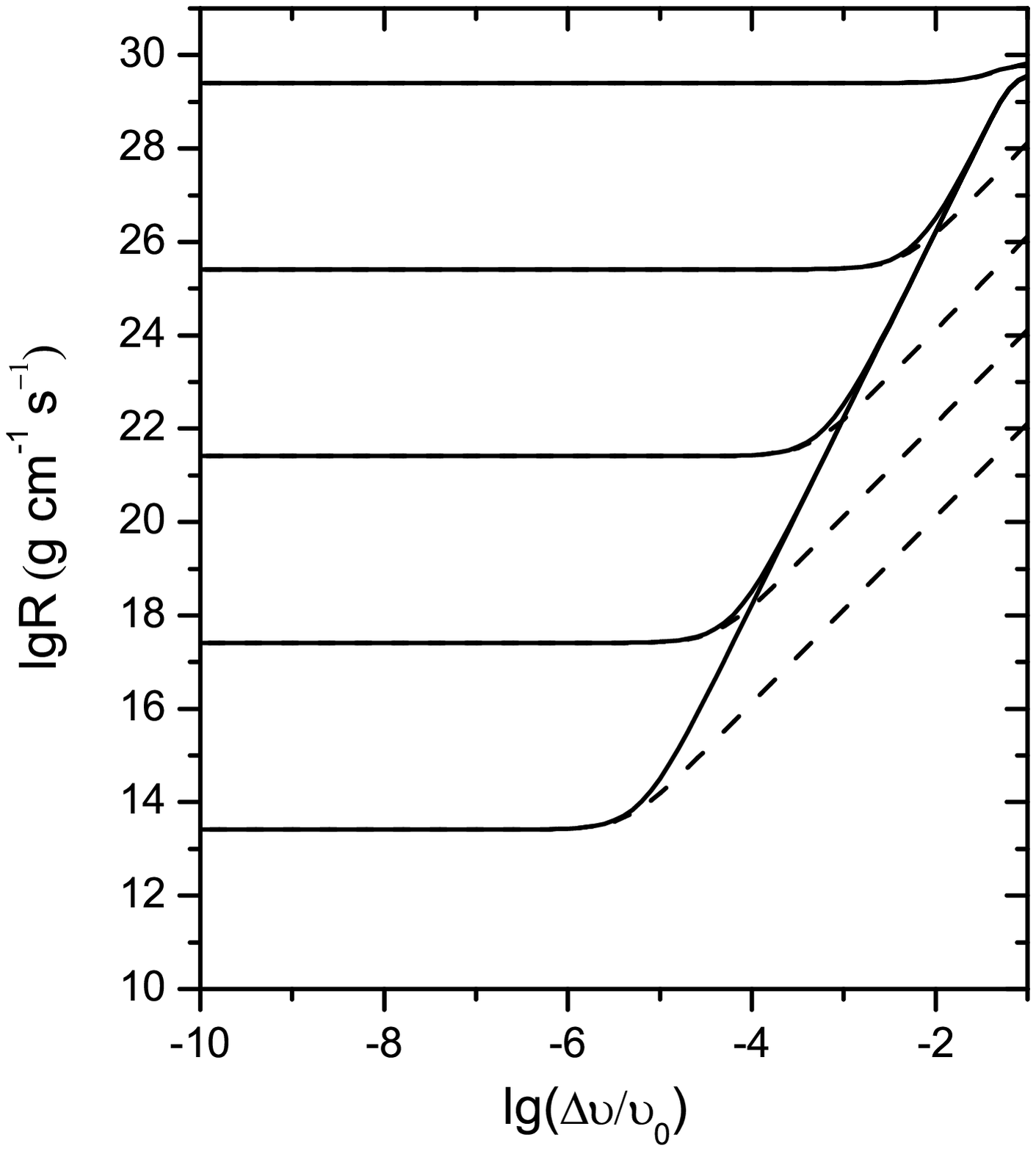}
   \caption{Radiative viscosity as function of relative
   volume perturbation amplitude for $\tau=10^{-3} \rm s$ and $\rho_{b}=0.6fm^{-3}$, where the direct Urca process occurs.
   The dash curves are the results up to $z^{4}$ and the solid curves up to $z^{6}$.
   The temperatures are $10^{-4}$, $10^{-3}$, $10^{-2}$, $10^{-1}$, and $1$ MeV  from bottom to top, respectively.}
   \label{Fig:f1}
   \end{figure}

 \begin{figure}
   \centering
   \includegraphics[width=0.5\textwidth]{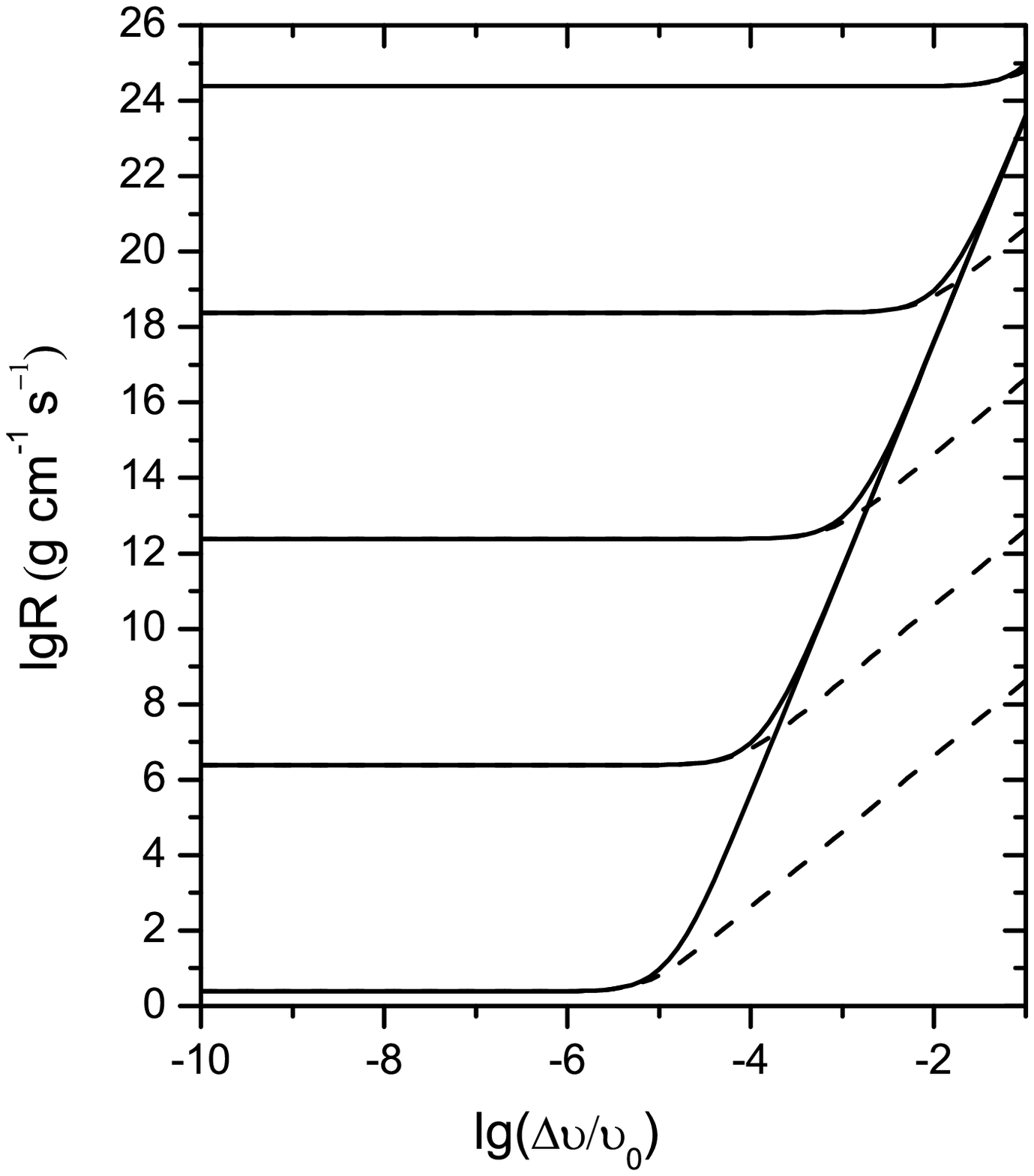}
   \caption{Radiative viscosity as function of relative
   volume perturbation amplitude for $\tau=10^{-3} \rm s$ and $\rho_{b}=0.3fm^{-3}$, where the direct Urca process is forbidden.
   The dash curves are the results up to $z^{4}$ and the solid curves up to $z^{8}$.
   The temperatures are $10^{-4}$, $10^{-3}$, $10^{-2}$, $10^{-1}$, and $1$ MeV  from bottom to top.
    }
 \label{Fig:f2}
 \end{figure}

\begin{figure}
\centering
\includegraphics[width=0.6\textwidth]{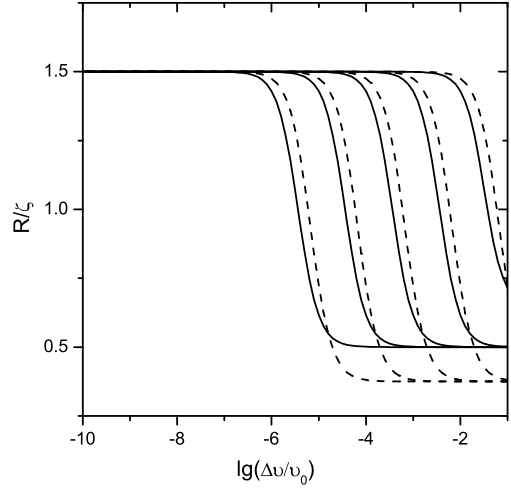}
\caption{${\cal R}/\zeta$  as function of relative
   volume perturbation amplitude for $\tau=10^{-3} \rm s$ , $\rho_{b}=0.6fm^{-3}$ (solid curves) and $\rho_{b}=0.3fm^{-3}$ (dash curves).
   The temperatures are $10^{-4}$, $10^{-3}$, $10^{-2}$, $10^{-1}$, and $1$  MeV from left to right. }
\label{Fig:f3}
\end{figure}

 \begin{figure}
   \centering
   \includegraphics[width=0.6\textwidth]{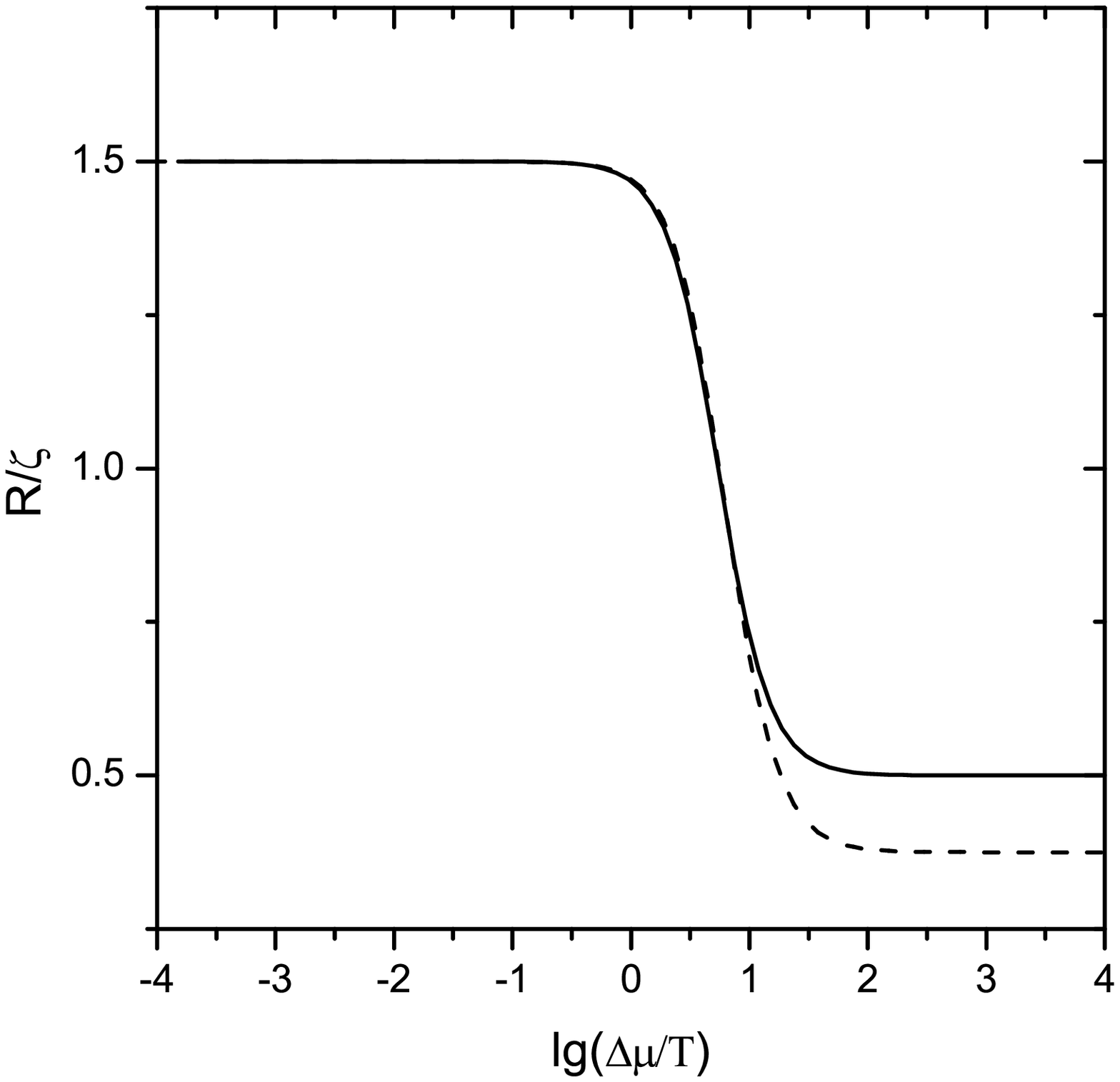}
   \caption{${\cal R}/\zeta$  as function of ${\rm lg} (\Delta\mu/T) $ for $\tau=10^{-3} \rm s$ , $\rho_{b}=0.6fm^{-3}$ (solid curves) and $\rho_{b}=0.3fm^{-3}$ (dash curves), where $\Delta\mu$ is the perturbation amplitude of chemical potential difference $\delta\mu(t)$.
    }
 \label{Fig:f4}
\end{figure}


\begin{thebibliography}{999}
\bibitem{mig79}A.B. Migdal, A.I. Chernoutsan, I.N. Mishustin, Phy. Lett. B 83 (1979) 158.

\bibitem{and98}N. Andersson, Astrophys. J. 502 (1998) 708.
\bibitem{fri98}J.L. Friedman, S.M. Morsink, Astrophys. J. 502 (1998) 714.
\bibitem{lin98}L. Lindblom, B.J. Owen, S.M. Morsink, Phys. Rev. Lett. 80 (1998) 4843.
\bibitem{mad00}J. Madsen, Phys. Rev. Lett. 85 (2000) 10.
\bibitem{and01}N. Andersson, K.D. Kokkotas, Int. J. Mod. Phys. D 10 (2001) 381.

\bibitem{fin68}A. Finzi, R.A. Wolf, Astrophys. J. 153 (1968) 835.
\bibitem{saw8901}R.F. Sawyer,  Phys. Rev. D 39 3804 (1989).
\bibitem{haensel92}P. Haensel, R. Schaeffer, Phys. Rev. D 45 (1992) 4708.
\bibitem{hae01}P. Haensel, K.P. Levenfish, D.G. Yakovlev,  Astron.\ Astrophys. 357 (2000) 1157.
\bibitem{wan84}Q.D. Wang, T. Lu, Phy. Lett. B 148 (1984) 211.
\bibitem{saw8902}R.F. Sawyer, Phy. Lett. B 233 (1989) 412.
\bibitem{mad92}J. Madsen, Phys. Rev. D 46 (1992) 3290.
\bibitem{gup97}V.K. Gupta, A. Wadhwa, S. Singh, J.D. Anand, Pramana - J. Phys. 49 (1997) 443.
\bibitem{lin02}L. Lindblom, B.J. Owen, Phys. Rev. D 65 (2002) 063006.
\bibitem{zhe02}X.P. Zheng, S.H. Yang, J.R. Li, Phys. Lett. B 548 (2002) 29.
\bibitem{zhe04}X.P. Zheng, X.W. Liu, M. Kang, S.H. Yang, Phys. Rev. C 70 (2004) 015803.
\bibitem{zhe05}X.P. Zheng, M. Kang, X.W. Liu, S.H. Yang, Phys. Rev. C 72 (2005) 025809.
\bibitem{pan06}N.N. Pan, X.P. Zheng, J.R. Li, Mon. Not. Roy. Astron. Soc. 371 (2006) 135.
\bibitem{sad0701}B.A. Sa'd, I.A. Shovkovy, D.H. Rischke,  Phys. Rev. D 75 (2007) 065016.
\bibitem{sad0702}B.A. Sa'd, I.A. Shovkovy, D.H. Rischke,  Phys. Rev. D 75 (2007) 125004.
\bibitem{don07}H. Dong, N. Su, Q. Wang, J. Phys. G 34 (2007) S643
\bibitem{sad09}B.A. Sa'd, J. Schaffner-Bielich, arXiv:0908.4190 [astro-ph].
\bibitem{lat91}J.M. Lattimer, C.J. Pethick, M. Prakash, P. Haensel , Phys. Rev. Lett. 66 (1991) 2701.
\bibitem{rei95}A. Reisenegger, Astrophys. J. 442 (1995) 749.
\bibitem{hae92}P. Haensel, Astron.\ Astrophys. 262 (1992) 131.
\bibitem{pra88}M. Prakash, T.L. Ainsworth, J. M. Lattimer, Phys. Rev. Lett. 61 (1988) 2518.
\end{thebibliography}
\end{document}